%Paper: astro-ph/9412049
%From: bp@astro.Princeton.EDU
%Date: Wed, 14 Dec 94 16:50:43 EST

%TeX MACRO for DARK MATTER conference proceedings
\magnification=\magstep1
\hsize=5.75truein
\vsize=8.75truein
\baselineskip=12.045truept % sets baseline spacing to 6 lines per inch %
\parindent=2.5em     % sets paragraph indent to 5 spaces %
\parskip=0pt
\def\hi{\noindent \hangindent=2.5em}
\def\bigskip{\vskip 12.045pt}  % skip one line %
\overfullrule=0pt
\font\twelverm=cmr10 at 12truept  % sets type size to 12 pt %
\twelverm    % document font 12 point Roman %
\nopagenumbers
\def\title#1{\centerline{\bf #1}}
\def\author#1{\bigskip\centerline{#1}}
\def\address#1{\centerline{#1}}
\def\sec#1{\bigskip\centerline{#1}\bigskip}
%%%%%%%%%%%%%%%%%%%%%%%%%%%%%%%%%%%%%%%%%%%%%%%%%%%%%%%%%%%%%%%%%%%%%%%%%%%%
% some useful but optional definitions
%%%%%%%%%%%%%%%%%%%%%%%%%%%%%%%%%%%%%%%%%%%%%%%%%%%%%%%%%%%%%%%%%%%%%%%%%%%%

\def\go{
\mathrel{\raise.3ex\hbox{$>$}\mkern-14mu\lower0.6ex\hbox{$\sim$}}
}
\def\lo{
\mathrel{\raise.3ex\hbox{$<$}\mkern-14mu\lower0.6ex\hbox{$\sim$}}
}
%%%%%%%%%%%%%%%%%%%%%%%%%%%%%%%%%%%%%%%%%%%%%%%%%%%%%%%%%%%%%%%%%%%%%%%%%%%%

{}~~~
\bigskip
\title{DARK MATTER -- PERSONAL VIEW}
\author{Bohdan Paczy\'nski}
\address{Princeton University Observatory, Princeton, NJ 08544-1001}
\address{bp@astro.princeton.edu}

\sec{ABSTRACT}

Traditional evidence for large amount of dark matter is based on
dynamical consideration for systems with $ t_{dyn} \gg t_{obs} $.
Recent observational and theoretical developments in gravitational
lensing offer a much more robust determination of the mass distribution
in some galaxies and their clusters, with the precision comparable
to that obtainable for double stars for which $ t_{dyn} < t_{obs} $,
and offer independent and direct evidence for the presence of dark matter.

Gravitational microlensing and femtolensing offer a possibility
to detect MACHOs with masses in excess of $ \sim 10^{-15} ~ M_{\odot} $.
The recent detections of microlensing events by the EROS, MACHO and
OGLE teams do not require any dark lenses as ordinary low mass stars
are compatible with the observations.  However, these searches
will soon either detect genuine MACHOs, or they will place stringent
upper limits on their number density.

\sec{1. INTRODUCTION}

The presence of dark matter is inferred through its gravitational
effect on the luminous matter.  The same basic concept led John C.
Adams and Urbain Le Verrier to the prediction of the existence of
a dark object which was responsible, through its gravity, for the
perturbations of the orbit of Uranus.  The object was discovered
by Johann Galle in 1946, and was named Neptune.  More recently accurate
masses of the dark matter binary system PSR 1913+16 were determined on
the basis of dynamical considerations (Taylor \& Weisberg 1989, and
references therein).  For Neptune and for PSR 1913+16 the mass to light
ratio (M/L) is enormous, much higher than for any galaxy or a cluster of
galaxies.  Presumably, these two types of dark objects, and their M/L,
are not typical for the universe at large.

Less precise dynamical consideration are the basis for the claim that
there is plenty of dark matter in clusters of galaxies (Zwicky 1933)
as well as in galaxies (Oort 1932, Freeman 1970).  The propagation of
light from distant sources is a very powerful tracer of mass distribution
through the phenomenon called gravitational lensing.

Various inferences have a different degree of reliability.  My
personal rating of the various methods is as follows:

\hskip 1.0cm {1) dynamical, with $ t_{dyn} < t_{obs} $ \hskip 1.2cm
-- the best } \hfill

\hskip 1.0cm {2) gravitational lensing \hskip 2.27cm -- good } \hfill

\hskip 1.0cm {3) dynamical, with $ t_{dyn} > t_{obs} $ \hskip 1.2cm
-- fair } \hfill

\hskip 1.0cm {4) mass to light ratio \hskip 2.55cm  -- poor } \hfill

\hskip 1.0cm {5) cosmological principles \hskip 1.89cm -- meaningless } \hfill

\noindent
where $ t_{dyn} $ is the dynamical time scale for the system, like
the binary period, or the virialization time, and $ t_{obs} $ is
the length of time over which the observations were carried out.

There is no doubt that the first technique provides a very robust mass
measurements.  This is the only truly fundamental method of measuring
masses in astronomy.  In practice it is applicable
only to binary stars and to planets.
The third method, with $ t_{dyn} > t_{obs} $,
or rather $ t_{dyn} \gg t_{obs} $, is by far the most commonly used
in modern inferences for the presence of dark matter.  Its reliability
varies a lot from case to case.
The second method, which only recently developed
to the extent that it is practical under a range of conditions,
is the main one to be discussed in this review.  It is very
direct, using light rays as tracers of space-time geometry, and it does
not require the mass distribution to be in any kind of equilibrium.  Its
main practical weakness is that in most cases the model of mass distribution
is not unique.  Fortunately, there are many cases in which models are
reasonably accurate, and the observational as well as modeling techniques
are developing rapidly.

The last two methods are far less reliable.  For some astronomical
object, be it a galaxy or a cluster of galaxies, the mass is estimated
with the method (3).  The luminosity of the system is measured directly,
and the M/L is calculated.  The estimates of M/L are available
for many galaxies, groups and clusters of galaxies, and
some average (median?) value of M/L is guessed to be representative
for the whole universe.  This value being of the order of
$ \sim 300 ~ M_{\odot} / L_{\odot} $ implies that most matter in the
universe is dark.

It is important to remember that we have no fundamental theory
that would allow us to calculate the efficiency of star formation
from diffuse gas or the stellar mass function.  It is not known
how the efficiency of star formation and the shape
of the mass function depend on the
initial gas pressure, magnetic fields, chemical composition, ambient
radiation field, local energy density in cosmic rays, etc.  Our knowledge
in this field is empirical only, and even this is incomplete and often
unreliable.  Therefore, my personal rating of the estimates
of the amount of dark matter in the universe as based on the
method (4) is `poor'.

My personal rating of the method (5) is `meaningless' even though
there is a strong theoretical bias towards
cosmology with $ \Omega = 1 $.  The common justification
for this bias comes from the theory of inflation, which in its
original form `predicted' $ \Omega = 1 $.  However, the modern theory
of inflation can accommodate almost anything: certainly $ \Omega = 0.3 $
(Kamionkowski et al. 1994, Bucher et al. 1994),
and a `tilted' (i.e. non-Zeldovich) spectrum of primordial perturbations
(Adams et al. 1992)
are among the many `predictions'.  Therefore, even though the inflation
remains a wonderful concept, and even though in some distant future it
may become rigorous, it currently provides no realistic estimate for
the amout of dark matter.  It is listed as a method number 5 for historical
reasons only.

There is no hope to use method (1) on cosmological scale, as typically
we may have $ t_{dyn} \sim 10^8 $ years, or even more, orders of magnitude
above $ t_{obs} $.  Hence the method (3) is by far the most common.
However, just as on the stellar scale the method (1) is the bedrock
of all secondary methods, so the method (2) should become the bedrock
for mass estimates on the galactic scale, even though its range of
applicability is rather limited so far.  There are many excellent recent
reviews and books on gravitational lensing:
Blandford \& Narayan (1992), Refsdal \& Surdej (1994),
Schneider, Ehlers, \& Falco (1992), Surdej at al. (1993).
As all important references can be found in these reviews, and the
page limit imposed on this paper is very strict, I make no
attempt to provide a complete or even fair reference to all important
contributions to this rapidly growing field, and I apologize for
any discomfort this may create.

\sec{2. GRAVITATIONAL MACROLENSING}

Strong gravitational lensing makes multiple images of a single
source.  If the separation between the images is large enough
(arcseconds in optical domain, milli arcseconds in the radio)
so that they can be seen separately, the phenomenon is referred
to as macrolensing.  The gravitational lensing is called weak
if only one distorted image of a source is present.

The largest scale on which the strong gravitational lensing was detected
and on which robust quantitative models of the lens mass distribution
are available are clusters of galaxies, in which `luminous arcs'
were discovered a few years ago (Lynds \& Petrosian 1989, and references
therein).  Later, less spectacular but much more common `arclets'
were found in many clusters.  These are caused by the weak gravitational
lensing and are so numerous that they proved to be very useful for the
studies of mass distribution (Tyson et al. 1990, and references therein).
The `arcs' and the `arclets' are the highly distorted images of galaxies
which are at a larger redshift than the cluster itself, and so they
are easily recognized being much bluer than the cluster galaxies.

The models of cluster mass distributions will be discussed later during
this symposium, so I would like to point out only a few obvious results.
First, the `arcs' demonstrate that the density of matter increases
strongly towards cluster centers and exceed the critical value needed
for strong gravitational lensing.  In other words the core radii
of the mass distribution are much smaller than believed only a decade
ago.  Second, the column mass density within
a cluster can be measured with the lensing model while the surface
brightness can be measured directly.  Hence, the value of M/L can be
determined with a fairly high accuracy.

Many gravitational lenses are caused by galaxies, which form a double or
a quadruple image of a distant quasar.  The total mass contained between
the images can be measured with a very high precision (Kochanek 1991),
and so can be the light of the lensing galaxy, thereby allowing
a precise M/L estimate to be made.  In some cases a model gives
a column mass density along the lines of
sight towards the individual images (cf. Kochanek 1994).

An interesting case is the double quasar 0957+561 (Walsh et al. 1979),
which has a VLBI radio jet some 40 milli arcseconds long.  Two different
images of the jet are observed.  The matter along the two lines of sight
is mostly dark and the millilensing properties are are different.  Therefore,
a detailed comparison of the two radio images offers an opportunity
to either detect the presence of $ \sim 10^6 ~ M_{\odot} $ compact objects
(black holes? dark clusters?) or demonstrate that such objects do not
exist (Wambsganss \& Paczy\'nski 1992, Garrett et al. 1994).

\sec{3. GRAVITATIONAL MICROLENSING}

A lensing galaxy is made of stars and some dark matter.  We know
nothing about the form of dark matter, but the stars can be treated
as point masses.  Each image of a lensed quasar is seen through
the lensing galaxy, and in turn is lensed by the stars close to the
line of sight, which split each image into a number of sub-images.
The sub-images are so close to each other that they cannot be seen
separately and the phenomenon is
called gravitational microlensing.  It can be detected because the
relative motion of the lens, the source, and the observer, or the motions
within the lens make the combined intensity of all micro-images vary.
There are two extreme types of microlensing, corresponding to a
very low and large `optical depth', respectively.  In the very low optical
depth limit there is either no, or at most one star close to the
line of sight, and there are either one or just two bright micro-images.
In cass of a large optical depth there are many stars close to the
line of sight and many bright micro-images are formed.  The properties of
lensing in these two regimes are very different.

The optical depth to gravitational microlensing is very low when we look
at the stars in our galaxy, or at the stars in the Magellanic
Clouds.  Along the lines of sight towards these stars there may be
other stars, brown dwarfs, planets and black holes acting as
gravitational microlenses.  The probability
that the brightness of a star is magnified by more than $ \sim 0.3 $ mag
is very low, $ \sim (V_{rot}/c)^2 \sim 10^{-6} $, where $ V_{rot} $ is
the rotational velocity of our galaxy (Paczy\'nski 1986, 1991, Griest
1991, Griest et al. 1991).  Three groups are conducting the search
for such events: EROS, MACHO, and OGLE, and all three reported the
detection of candidate events towards the LMC and/or the galactic bulge
(Alcock et al. 1993, 1995, Aubourg et al. 1993, Udalski et al 1993,
Udalski et al. 1994a, 1994b, 1994c, Bennett 1994)

This is a field at its infancy, but some reasonably firm
conclusions are already available.  First, it has been demonstrated that
it is possible to detect the very rare events of gravitational microlensing
at a rate of a few per year (OGLE) or even a few dozens per year (MACHO).
Second, it is possible to detect the events in real time with the
OGLE `early warning system' EWS (Paczy\'nski 1994a, Udalski et al.
1994c), and with the MACHO `alert system' (Bennett et al. 1994).
This is important, as the follow-up observations are possible while the
event unfolds.  Third, the rate of events towards the galactic
bulge turned out to be higher than expected by a factor of 2-4,
indicating that our understanding of the galactic structure is
inadequate.  It follows, that microlensing offers a new way
to study the galactic structure.  Four, even though the search for
microlensing events was undertaken with the dark matter in mind,
the results available so far neither prove nor disprove the hypothesis
that dark matter is made of massive compact objects (MACHOs).  All
events detected so far are compatible (within observational and
theoretical errors) with the lenses being ordinary stars.
My personal conclusion is that MACHOs, if they exist, will be
detected within the next few years.

There is one major misconception which invalidates many theoretical
studies: the incompleteness of our knowledge about the galactic
structure means that currently there is no meaningful way to relate
the observed time scales of microlensing events to the lens masses,
as it is not clear if the lenses detected towards the galactic
bulge are in the galactic disk (Alcock et al. 1995), or are they
in the galactic bar (Paczy\'nski et al. 1994), while the lenses
detected towards the LMC may be in the disk or the halo of our galaxy,
or they may be in the bulge
or the halo of LMC.  The inferences about the lens masses will become
realistic when the location of lenses is established with the
future observations.

There are also misconceptions about the statistical properties
expected of microlensing events.  As the events are so rare it is
commonly believed that they should not repeat.  In fact we do know
that the stars are commonly double, and the same presumably holds for the
lenses.  Mao \& Paczy\'nski (1991) estimated that
$ \sim 10\% $ of all events may have a strong signature
of a double lens.  Indeed, in a small sample of 13 events detected by
the OGLE there is one very dramatic case of a double lens, the
OGLE \#7 (Udalski et al. 1994d, Bennett et al 1994), in which the two
lensing stars were separated by approximately one Einstein ring radius.
Another possible OGLE double lens was analyzed by Mao \& DiStefano (1995).
The distribution of binary separations is uniform in the logarithm
of separation.  Therefore, there should also be
sources microlensed by the two components of a binary
with the separation exceeding the Einstein ring radius.  In this
case there should be two separate single microlensing events,
separated by a few months or a few years.
A similar dual event might be observed if the source is double
(Griest 1992, Griest \& Hu 1993).  I rougly estimate that a few
percent of all events should be like that, i.e. they should repeat.
If such repeating cases are not found then either binary stars are less
common than we think, or the detection criteria discriminate against them.

A very different regime of microlensing and a very different type
of variability is expected when the optical depth is modest or even large.
This is the case of a quasar lensed by a galaxy
at a cosmological distance.  The first clear case of such microlensing
was reported for 2237+0305, i.e. Huchra's lens
(Huchra et al. 1985, Irwin et al. 1989).  The most dramatic event was
reported by Pen et al. (1994): the luminosity of one of the four quasar images
increased by 1.5 mag during a time interval of $ \sim 2 $ months, and
declined by the same amount during a few days.  This is compatible
with a theoretical picture in which the magnification pattern
is a maze of caustics
produced by the stars randomly distributed in the lensing galaxy
(Wambsganss et al. 1990).  The most common light variation is caused
by the source (the quasar) crossing one of the many caustics.  The
variation on one side is relatively slow, with the time scale proportional
to the square root of a typical stellar mass.  The variation on the
other side would be instantaneous for a point source, and has
a finite rate for an extended source, with the duration proportional
to the source size.  Still, there is a surprise in the Pen et al. (1994)
result: it implies the quasar is as small as $ \sim 10^{14} ~ cm $ in
the continuum light, much smaller than expected in the conventional
accretion disk models (Rauch \& Blandford 1991, Jaroszy\'nski et al. 1992).

There are also misconceptions about the microlensing in the large
optical depth regime.  Contrary to common belief it is not possible to
relate the observed light variation to a specific stars, as the
caustics are formed by many stars acting together,
gravity being a long range force,  Also, the relation between
the time scale of the light variability and the microlens masses has not
been worked out in any paper published so far.  Even though the time scale
has to be proportional to a
square root of some average microlens mass, the dimensionless coefficient
of proportionality is not the same as it is in the
optically thin case.  Therefore, it is not possible (at this time)
to estimate the microlens masses in the Huchra's lens to better than
an order of magnitude.

\sec{4. GRAVITATIONAL FEMTOLENSING}

Gravitational microlensing, as well as macrolensing, operates within
the approximation of geometrical optics.  Yet, under certain
conditions the diffraction effects might be important.  It is
required that the photon wavelength must be longer than the Schwarzschild
radius of the lensing mass, and the source has to be smaller than
the Fresnel length.  Gould (1992) was the first to point out that
this effect may be used to search for dark matter made of objects
of $ \sim 10^{-15} ~ M_{\odot} $, as they might show
up as anomalous spectral features in gamma-ray bursts at cosmological
distance.  Gould also invented the name `femtolensing' to describe the
phenomenon.  The theory was farther elaborated by Stanek et al. (1993)
and by Ulmer \& Goodman (1994).  Jaroszy\'nski (1994) found that
the diffraction effects can be present even when the photon wavelength
is much smaller than the lens Schwarzschil radius, and in fact they limit
the resolving power of caustics in Huchra's lens to $ \sim 10^{-12} $
seconds of arc.  No case of gravitational femtolensing has been detected
as yet.

\sec{5. THE VALUES OF $ H_0 $ , $ \Omega $ and $ \Lambda $ -- GRAVITATIONAL
MEGALENSING}

The three most important cosmological parameters: $ H_0 $, $ \Omega $, and
$ \Lambda $ may be determined using gravitational lensing effects.
When two well separated images of the source can be observed, and
the source is variable, then the time delay between the variations
as observed in the two images is proportional to the Hubble time, i.e.
to $ H_0^{-1} $ (Kayser 1993, and references therein).  The other
parameters may also be estimated using the lens statistics (King 1993,
and references therein).  However, the classical method of estimating
$ \Omega $ and $ \Lambda $ is based on the redshift - angular diameter,
or equivalently redshift - luminosity relation.  Fundamentaly this
is the phenomenon of gravitational lensing by the universe as
a whole, and it might be called megalensing.  Currently, the best prospect
to apply this classical method may be offered by supernovae of Type Ia
(Colgate 1979, Branch \& Tammann 1992, and references therein,
Phillips 1993).

\sec{6. CONCLUSIONS}

Strong gravitational macrolensing offers the most direct, and perhaps
the most accurate determination of the mass of cosmological objects like
galaxies and their clusters.  Unfortunately, this method is applicable
only to the objects with near critical column mass density.  The weak
gravitational macrolensing can be applied in more general, lower density
conditions, but it is also less direct and hence less reliable.  In
some cases it is possible to measure directly the column mass density
within the beam towards the images.  Gravitational macrolensing can be
used (and it is used) to establish the presence of dark matter of some kind,
though so far it does not distinguish between the MACHOs and the WIMPS.

Gravitational microlensing (and femtolensing) can provide the most direct
determination of the presence of MACHOs in a very large range of masses,
from $ 10^{-15} ~ M_{\odot} $ to $ 10^{6} ~ M_{\odot} $ (Nemiroff 1993,
and references therein), but so far there is no definite evidence for
or against MACHOs in any mass range.  However, the recent successes of
the MACHO and OGLE teams in detecting microlensing within our galaxy
suggest that genuine MACHOs will be detected within a few years.

It is possible that supernovae Type Ia will allow the determination
of $ \Omega $ and $ \Lambda $ using the classical redshift - luminosity
relations, which are related to gravitational lensing by all matter
within the beam of radiation - gravitational megalensing.  The Hubble
constant may be measured, at least in principle.
with the time delay effect.

Clearly, gravitational lensing is not the only way to study cosmology
and to search for dark matter.  However, it is one of the most useful
and versatile approaches, and in some cases (mass determination) it
is the most reliable.

\sec{ACKNOWLEDGEMENTS \& APOLOGIES}

It is a great pleasure to acknowledge that many of the ideas presented
in this review are the result of numerous conversations and discussions
with many friends, too many to list them all.  I apologize for
not being more specific.  I also apologize for not providing more
complete references and for not mentioning many important contributions.

This project was supported with the NSF grant AST 93-13620 and NASA
grant NAG5-1901.

\sec{REFERENCES}

\hi {Adams, F. C. et al. 1992, Phys. Rev. D, 47, 426}

\hi {Alcock, C., et al. 1993, Nature, 365, 621}

\hi {Alcock, C., et al. 1995, ApJ, in press}

\hi {Aubourg, E., et al. 1993, Nature, 365, 623}

\hi {Bennett, D. P. 1994, this conference}

\hi {Blandford, R. D., \& Narayan, R. 1992, AAR\&A, 30, 311}

\hi {Branch, D., \& Tammann, G. A. 1992, ARA\&A, 30, 359}

\hi {Bucher, M., Goldhaber, A. S., \& Turok, N. 1994, preprint
iassns-hep-94-81, PUPT-94-1507}

\hi {Colgate, S. A. 1979, ApJ, 232, 404}

\hi {Freeman, K. C. 1970, ApJ, 161, 802}

\hi {Garrett, M. A. et al. 1994, MNRAS, 270, 457}

\hi {Griest, K. 1991, ApJ, 366, 412}

\hi {Griest, K., et al. 1991, ApJ, 372, L79}

\hi {Griest, K. 1992, ApJ, 397, 362}

\hi {Griest, K., \& Hu, W. 1993, ApJ, 407, 440}

\hi {Huchra, J., et al 1985, AJ, 90, 691}

\hi {Irwin, M. J., et al. 1989, AJ, 98, 1989}

\hi {Jaroszy\'nski, M., Wambsganss, J., \& Paczy\'nski, B.
1992, ApJ, 396, L65}

\hi {Jaroszy\'nski, M. 1994, in preparation}

\hi {Kamionkowski, M., Ratra, B., Spergel, D. N., \& Sugiyama, N. 1994,
ApJ, 434, L1}

\hi {Kayser, R. 1993, in Gravitational Lenses in the Universe, p. 5,
Surdej, J. at al. eds., Proc. 31st Li\`ege Coll.}

\hi {King, P. 1993, in Gravitational Lenses in the Universe, p. 53,
Surdej, J. at al. eds., Proc. 31st Li\`ege Coll.}

\hi {Kochanek, C. S. 1991, ApJ, 373, 354}

\hi {Kochanek, C. S. 1991, ApJ, submitted = Harvard - Smithsonian CfA
Preprint No. 3933}

\hi {Lynds, R., \& Petrosian, V. 1989, ApJ, 336, 1}

\hi {Mao, S., \& Paczy\'nski, B. 1991, ApJ, 374, L37}

\hi {Mao, S., \& DiStefano, R. 1995, ApJ, in press}

\hi {Nemiroff, 1993, in Gravitational Lenses in the Universe, p. 53,
Surdej, J. at al. eds., Proc. 31st Li\`ege Coll.}

\hi {Oort, J. H. 1932, BAN, 6, 249}

\hi {Paczy\'nski, B. 1986, ApJ, 304, 1}

\hi {Paczy\'nski, B. 1991, ApJ, 371, L63}

\hi {Paczy\'nski, B. 1994a, IAU Circ. No. 5997}

\hi {Paczy\'nski, B., et al. 1994, ApJ, 435, L113}

\hi {Pen, Ue-Li, et al. 1993, in Gravitational Lenses in the Universe,
p. 111, Surdej, J. at al. eds., Proc. 31st Li\`ege Coll.}

\hi {Phillips, M. M. 1993, ApJ, 413, L105}

\hi{Rauch, K. P., \& Blandford, R. D. 1991, ApJ, 381, L39}

\hi {Refsdal, S. \& Surdej, J. 1994, Rep. Prog. Phys., 56, 117}

\hi {Schneider, P., Ehlers, J., \& Falco, E. E. 1992, Gravitational
Lenses (Springer Verlag)}

\hi {Stanek, K. Z., Paczy\'nski, B., \& Goodman, J. 1993, ApJ, 413, L7}

\hi {Surdej, J. at al. eds. 1993, Gravitational Lenses in the Universe,
Proc. 31st Li\`ege Coll.}

\hi {Taylor, J. H., \& Weisberg, J. M. 1989, ApJ, 345, 434}

\hi {Tyson, J. A., Valdes, F., \& Wenk, R. A. 1990, ApJ, 349, L1}

\hi {Udalski, A., et al. 1994a, ApJ, 426, L69}

\hi {Udalski, A., et al. 1994b, Acta Astron., 44, 165}

\hi {Udalski, A., et al. 1994c, Acta Astron., 44, 227}

\hi {Udalski, A., et al. 1994d, ApJ, 436, L103}

\hi {Ulmer, A., \& Goodman, J. 1994, ApJ, submitted = Princeton Observatory
Preprint No. 569}

\hi {Walsh, D., Carswell, R. F., \& Weyman, R. J. 1979, Nature, 279, 381}

\hi {Wambsganss, J., \& Paczy\'nski, B. 1992, ApJ, 397, L1}

\hi {Wambsganss, J., Paczy\'nski, B., \& Schneider, P. 1990, ApJ, 358, L33}

\hi {Zwicky, Z. 1933, Helv. Phys. Acta, 6, 110}

\vskip 3.0cm

This is to appear in the proceedings of the 5th Annual October Astrophysics
Conference in Maryland: DARK MATTER, which took place on October 10-12,
1994 in College Park, Maryland.

\vfill
\bye